\documentclass[a4paper]{article}
\usepackage{natbib}
\usepackage{tgpagella}
\usepackage{geometry}
\usepackage[none]{hyphenat}
\title{Tracing Networks of Knowledge in the Digital Age}
\author{Mirco Musolesi \\ University College London and The Alan Turing Institute}
\date{}

\begin{document}
\sloppy
\maketitle

\begin{abstract}
The emergence of new digital technologies has allowed the study of human behaviour at a scale and at level of granularity that were unthinkable just a decade ago. In particular, by analysing the digital traces left by people in their online and offline lives, we are able to trace the spreading of knowledge and ideas at both local and global scales.

In this article we will discuss how these digital traces can be used to map knowledge in online and offline worlds, outlining both the limitations and the challenges in performing this type of analysis. We will focus on data collected from social media platforms, large-scale digital repositories and mobile data. Finally, we will provide an overview of the tools that are available to scholars and practitioners for understanding these processes using these emerging forms of data. 

\end{abstract}

\bigskip

\section{Overview}
Thanks to the existing and emerging digital technologies, nowadays it is possible to trace networks of knowledge at a scale and at a level of granularity that were unimaginable just a few years ago. In this article we will discuss a series of studies concerning the analysis and mapping of networks of knowledge at different scales through a variety of sources of digital data. We will focus in particular on data from social media, large-scale digital repositories and mobile data.

For instance, networks of knowledge can be extracted by analysing the interactions happening in social network platforms. In particular, models of information diffusion can be defined to represent and understand the spreading of knowledge in groups of users in time and space~\citep{nekovee2007theory, liben2008tracing,  lazer2009life, lerman2010information, kitsak2010identification, dodds2011temporal, karsai2011small, pastor2015epidemic}. A typical way is to describe the process of dissemination of the information using epidemic models, where the process of diffusion does not involve a disease~\citep{anderson1992infectious}, but a piece of information. In this way, it is possible to represent knowledge dissemination over networks as epidemic processes and formalise them by means of mathematical models~\citep{keeling2005networks}. One of the most fascinating aspects related to tracing networks of knowledge in the digital world is that it is possible to extract data in a digital format and use them to develop and test models of dissemination at a scale. Indeed, this is one of the most compelling examples of applications of big data outside the commercial world~\citep{musolesi2014big}.

Diffusion models can be devised to understand how knowledge `spreads' through a network of people in a community, in a society, in a city, in a country or even at planetary scale. Knowledge is usually defined as information acquired through experience or education. Therefore, tracing the diffusion of knowledge corresponds in a sense to tracing the diffusion of information~\citep{hidalgo2015information}. We have witnessed the development of a variety of models for describing the spreading and the adoption of knowledge, such as, for example, the generation and acceptance of innovative ideas, starting from the marketing community~\citep{goldenberg2001talk}. Network scientists have also contributed to the fields through the definition of several models of increasing complexity~\citep{iacopini2018network}. One of the most famous models is indeed the Bass model~\citep{Bass01011969}, which describes the adoption of innovation in a community by means of a very parsimonious mathematical model.

An interesting class of emerging models is that describing dynamical~\citep{barrat2008dynamical}, temporal~\citep{tang2010small, holme2012temporal} and spatio-temporal networks~\citep{Williams160196}. These models can be used for example to identify influencers and mediators in time-varying social networks~\citep{tang2010analysing}. Spatio-temporal network models can also be used to represent the spreading of information in time and space through networks of people (or places) and study its dynamics.

\section{Social media}

Social media can be used to track the spreading of information as it happens in real-time at a very large scale. They are extremely popular, involving thousands of millions of users.  The first example of information diffusion in the digital world indeed happens through Web blogs, which predate the advent of social media outlets such Facebook~\citep{ellison2007benefits} and Twitter~\citep{kwak2010twitter,cha2010measuring}. One of the first study of information diffusion in these media can be found in~\citep{gruhl2004information,adar2005tracking}. The geography of Twitter has been analysed in several works, such as~\citep{leetaru2013mapping}. Locative media has been the focus of investigation of several researchers, focussing on the interplay between our online and offline lives~\citep{ozkul2013you,ozkul2015record,ozkul2017mobile}. More in general, social media can contribute substantially to opinion formation, as studied for example in~\citep{watts2007influentials}.

In~\citep{srep02980} the authors examine the spreading of rumours related to the discovery of the Higgs boson~\citep{Aad20121,Chatrchyan201230} in real-time from Twitter data. Twitter is a social network where a user can write microblog posts composed of 140 characters. A post can then be \textit{retwitted} (i.e., reposted) by other users that are following a person or an organisation (\textit{followers}). The availability of data from the platform also allows for the extraction of mathematical models of diffusion in the social network of followers across the globe. Furthermore, the study of the topics discussed in Twitter has attracted the attention of a vast number of researchers in the past years. For example, in~\citep{romero2011differences} the authors analyse the adoption of hashtags and they show that for example those that are politically controversial are particularly persistent in the network over time.

Another interesting example of tracing knowledge using digital data is indeed the analysis of software developers' activities in GitHub, an online software development collaborative tool. In~\citep{Lima14:coding} the authors examine the network of collaborations on different software projects at a global scale, looking at how collaborations unfold, including their geographic characteristics. In GitHub users can create code repositories; every repository has a list of collaborators, who can make changes to the content of the repository. A user can submit changes to the codebase or if they do not want to be a collaborator, there is the possibility of `forking' a project. The action of fork creates a duplicate of the repository for independent work. By analysing the interactions on GitHub, it is possible to observe patterns of collaborations over time and over space. Indeed, software production is a creative process: collaboration on software projects can be seen as a new way of sharing knowledge and participating to the co-creation of intellectual artefacts.

One of the limitations of this body of work is the fact that often external factors cannot be captured, such as for example the influence of other social media or press outlets, e.g., television and newspapers and other online and offline media. An interesting study about evaluating the external influence of other sources on information spreading can be found in~\citep{myers2012information}.

The diffusion of social media has also led scholars to ask questions related to the importance of distance not only in a world connected by global online networks~\citep{mok2010does}, but also at city level~\citep{hristova16:multilayer}. In~\citep{scellato2010distance}, by analysing four social networks (Brightkite, Foursquare, LiveJournal and Twitter), the authors show that distance still matters in the establishment of social links: we cannot talk about \textit{death of distance}~\citep{cairncross2001death}, at least yet, in the case of social media. The resulting geo-social networks show heavy-tailed degree distributions, a tendency to exhibit users with high node locality and social triangles on a local geographic scale. Hristova et al. showed that geo-social networks in different platforms are also highly correlated~\citep{hristova16:multilayer} and this fact can be used for  effective link prediction.

Other interesting problems include the identification of influential spreaders of information~\citep{kitsak2010identification}, the maximisation of the spreading process itself~\citep{kempe2003maximizing}, and the impact of homophily in the diffusion of information~\citep{aral2009distinguishing}. Thanks to the availability of geo-social networks, it is also possible to identify the key spreaders in space~\citep{lima2012spatial}: this allows us to understand how specific people in a spatial network contribute to the dissemination of information and ideas in certain geographic areas, such as neighbourhood, cities or states. It is worth noting that the study of information diffusion in social media has not been limited to text but also focussed on other multimedia content such as photos in Flickr~\citep{cha2009measurement}. It has been argued that digital systems are defining the space itself and are changing the way we live our everyday lives~\citep{kitchin2011code}. At the same time this rich set of information is collected and can be analyzed, also in real-time, reconstructing the dense fabric of interactions of millions of individuals.

Information can also be political in nature such as in the case of protests~\citep{gonzalez2011dynamics, gonzalez2013broadcasters, margetts2015political}. For example, by retrieving information through social media, it is possible to map the evolution of protests, including the exchange of information, of groups operating at local and global level. This offers an unprecedented opportunity for studying and understanding political phenomena in real-time. Another very interesting aspect is that, for the first time, researchers are able to follow the spreading of information involving single actors and their thoughts at a very fine-grained granularity both in space and time. Moreover, it is worth noting that this information is not mediated through recollection, but it is instantaneous in nature, i.e., it is generated and can be collected as events unfold.

It is worth noting that social media data have been used for a very large number of studies, for example for understanding not only the spreading of information itself, but also the inherent contextual factors influence it, such as demographics and culture.
For example, one of the topics of interest of social media studies has been the analysis of surnames, which can be used to understand the geo-demographics of different areas of large cities, such as London~\citep{longley2016geo}. In this way, it is possible to trace back global networks of people and migrations reflected in the ethnic composition of large-scale cities. Geo-social networks can also be used effectively to trace and study diversity and gentrification phenomena~\citep{hristova16:measuring}. Finally, in~\citep{Silva14:youare} the authors discuss how data from the Foursquare platform can be used to understand cultural characteristics of geographic areas by looking at the presence of specific food and drinks venues.

\section{Large-scale data repositories}

Wikipedia, an online encyclopaedia that is completely edited by volunteers, is probably the most successful example of crowdsourced work. Through the analysis of Wikipedia usage in different languages and the application of data mining techniques to the content of the various articles, it is possible to reconstruct how different topics are of interest and relevant in different part of the globe~\citep{brandes2009network,yasseri2012dynamics}. Moreover, the articles themselves can be geo-located, i.e., mapped to different areas of the planet. In other words, it is truly possible to create maps of knowledge that connect people, objects, facts and places around the world. In order to extract information from the text usually natural language techniques (NLP) techniques~\citep{manning1999foundations} are applied. With respect to the extraction of geographic information, geo-localisation techniques are used to map names to geographic named entities. The mapping is performed by trying to map names to geographic locations that are available in geographic gazetteers such as Geonames~\citep{ahlers2013assessment}.

As a more general trend, we observe that citizens are getting more and more involved in the creation of knowledge, both contributing through text and multimedia content (images, videos, etc.) but also by acting as sensors and collecting information about the environments where they live~\citep{goodchild2007citizens}.  One of the most interesting examples is OpenStreetMap~\citep{openstreetmap}. Through the collaboration of a large number of individuals, knowledge is created, shared and disseminated. Several researchers have been investigated how this collaboration takes place, characterising his dynamics~\citep{hristova2013life}.

Other repositories that allow for the analysis of networks of knowledge include online literary repositories, such as for example Gutenberg~\citep{gutenbergproject}. The analysis of the networks emerging in the literature have been a theme of the work of Moretti among the others~\citep{moretti2005graphs}. Indeed, one possibility is to try to extract names of geographic locations as discussed above for Wikipedia from text in order to reconstruct how literary works are mapped in the geographic space, in order to understand how the fictional worlds are related to the real ones.

Another interesting example is the tracing of influence through citation networks~\citep{bilke2001topological}. These can be traced over both time and space and they provide a way for mapping out collaborations among individuals in different countries in a very detailed way considering exchange information across disciplines~\citep{porter2009science,sinatra2016quantifying} and in specific communities~\citep{lehmann2003citation}. An entire new academic discipline, usually referred to as ``science of science"~\citep{fortunato2018science}, focussing on understanding the dynamics of scientific collaboration and exchange of ideas is emerging.

\section{Mobility data}

Another interesting source of data is related to the increasing availability of mobility traces datasets, from flights~\citep{guimera2005worldwide} to migration patterns~\citep{davis2013global} data, from social media~\citep{noulas2012tale,hawelka2014geo} to mobile data collected directly by means of users' smartphones~\citep{internetcomputing,srep09136,CanMus15_trajectories} or by the cellular phone operators through the network infrastructure~\citep{laurila2012mobile}.

In this case, the network of knowledge emerging from the data is in a sense indirect, but through this information we can map how people move at different geographic scale and spread ideas. Indeed, as discussed above, some people have talked about the death of distance~\citep{cairncross2001death} as a consequence of the emergence of digital technologies, including for example the use of smartphones and global network connectivity. However, this claim is disputed by other researchers that are stressing the fact that distance is still a major factor in human relationships and, therefore, in spreading of ideas~\citep{scellato2010distance}.

In general, one of the most difficult aspects of using mobility data for understanding spreading processes is the fact that data provide only a partial view of the interactions between people. Indeed, it is extremely difficult to collect ground-truth information for reconstructing interactions from these datasets. For these reasons, it is always very important to be extremely cautious in deriving universal models from these datasets in terms of human behaviour and interactions.

\section{Tools}
In this section, we will briefly review the tools that are used for tracing networks from digital data. We can broadly divide them into three main categories: data collection, data analysis and data visualisation.

First of all, \textit{data collection} is a fundamental phase in the process of tracing networks of knowledge. Data is usually collected from digital sources such as the Web through scraping tools~\citep{mitchell2015web} or from digital platforms by means of Application Programming Interfaces (APIs) such as those provided by Twitter~\citep{twitterAPI}. A limited amount of data is usually available for free, such as in the case of Twitter.

Secondly, \textit{data analysis} is performed with a variety of tools from stand-alone applications such as Gephi~\citep{gephi} to purpose-built software developed using existing libraries such as NetworkX (for Python)~\citep{networkx} or igraph (available for a variety of programming languages)~\citep{igraph}. There are also several libraries for spatial analysis, such as PySAL~\citep{pysal} for Python.

Finally, as far as \textit{data visualisation} is concerned, various tools can be used for representing the networks of knowledge extracted from the data at local, national and global scales. An emerging class of tools is represented by open source tools for visualisation. Examples include again Gephi~\citep{gephi} or Cytoscape~\citep{cytoscape}. Network data visualisation is also performed through tools developed by means of `general-purpose' programming languages and platforms for data manipulation and visualisation such as Processing~\citep{reas2007processing}. Other data visualisation tools include classic Geographic Information Systems~\citep{longley2015geographic}, which provide functionalities for mapping networks in the geographic space and more advanced digital cartography visualisations~\citep{cheshire2014}.

\section{Outlook}

In this paper, we have examined new emerging and fascinating ways of tracing networks of knowledge in real-time using the digital traces that we leave in our everyday lives. We have explored the new possibilities offered by the availability of a variety of new forms of data, including social media, large-scale data repositories and mobile data. These data sources allow for new types of investigation that were not possible just a few years ago. At the same time, it is important to consider the limitations that are inherent in using these new forms of data for mapping social phenomena. One of the key issues is indeed associated to the intrinsic bias that is associated to them For example, the demographics of social media users might not be representative of the general population~\citep{mislove2011understanding}. However, it is worth noting that these sources are still valuable for understanding the behaviour of particular subsets of the population. In general, it is still possible to derive population-level observations taking into consideration at the same time the limitations of the data sources used for the analysis. Predictive models using techniques from Machine Learning and Artificial Intelligence are also opening up new possibilities in this field. For example, a very interesting and promising area is the creation of knowledge graphs, linking concepts, people and places~\citep{nickel2015review}. The application of machine learning to the process of understanding processes of knowledge dissemination and its implications has also recently attracted the attention of social scientists (see for example~\citep{mackenzie2017machine}, in which the author develops an archaeology of machine learning operations following Foucault~\citep{foucault2013archaeology}).

Other important considerations are related to the potential privacy issues associated with these new emerging sources of data~\citep{musolesi2014big,RosMus15_spatiotemporal}. It is worth noting that most of the data cited in this article are the results of aggregation, i.e., they preserve the identity of the single users. 

Most of the existing work has been based on extracting, mapping and visualising networks of knowledge. We believe that modelling and predicting the dissemination of knowledge represent a very interesting and still open areas for researchers, also considering the recent results in both network science~\citep{boccaletti2006complex, barrat2008dynamical, newman2010networks,pastor2015epidemic} and machine learning~\citep{Mur12_Machine,goodfellow2016deep}. This is indeed an area where truly interdisciplinary work is essential given the fact that the research questions in this field are really at the interface of the humanities, social sciences, computational and mathematical sciences.

\bigskip

\noindent{\bf Biographical Note}

\noindent Mirco Musolesi is Professor in Data Science at the Department of Geography at University College London and Turing Fellow at the Alan Turing Institute. At UCL he leads the Intelligent Social Systems Lab. He received a PhD in Computer Science from University College London and a Master in Electronic Engineering from the University of Bologna. After postdoctoral work at Dartmouth College and Cambridge, he held academic posts at St Andrews and Birmingham. The research focus of his lab is on sensing, modelling, understanding and predicting human behaviour in space and time, at different scales, using the `digital traces' we generate daily in our online and offline lives. He is interested in developing mathematical and computational models as well as implementing real-world systems based on them. This work has applications in a variety of domains, such as ubiquitous and autonomous systems design, healthcare and security\&privacy.

\bibliographystyle{agsm}
\bibliography{biblio}

@inproceedings{Lima14:coding,
 author = {Antonio Lima and Luca Rossi and Mirco Musolesi},
 title = {{Coding Together at Scale: GitHub as a Collaborative Social Network}},
 booktitle = {Proceedings of ICWSM'14},
 address = {Ann Arbor, Michigan, USA},
 month = {June},
 year = {2014}
}

@misc{twitterAPI,
 title={Twitter REST API},
 note={https://developer.twitter.com/en/docs [Accessed on 7 August 2019]},
 year={2019}
 }

@misc{networkx,
 title={Networkx Library},
 note={https://networkx.github.io [Accessed on 7 August 2019]},
 year={2019}
 }

@misc{gutenbergproject,
 title={The Gutenberg project},
 note={https://www.gutenberg.org [Accessed on 7 August 2019]},
 year={2019}
 }

@misc{openstreetmap,
 title={OpenStreetMap},
 note={http://www.openstreetmap.org [Accessed on 7 August 2019]},
 year={2019}
 }

@misc{igraph,
 title={igraph},
 note={http://igraph.org/ [Accessed on 7 August 2019]},
 year={2019}
 }

@misc{pysal,
  title={PySAL},
  note={http://pysal.org},
  year={2019}
  }

@misc{gephi,
 title={Gephi - The Open Graph Viz Platform},
 note={c [Accessed on 7 August 2019]},
 year={2019}
 }

@misc{cytoscape,
 title={Cytoscape: An Open Source Platform for Complex Network Analysis},
 note={http://www.cytoscape.org [Accessed on 7 August 2019]},
 year={2019}
}

@article{srep02980,
	author = {Manlio {De Domenico} and Antonio Lima and Paul Mougel and Mirco Musolesi},
	title = {The Anatomy of a Scientific Rumor},
	journal = {Scientific Reports},
	volume = {3},
	month = {October},
	year = {2013},
	publisher = {Nature Publishing Group},
	number = {02980}
}

@inproceedings{myers2012information,
  title={Information diffusion and external influence in networks},
  author={Myers, S.A. and Zhu, C. and Leskovec, J.},
  booktitle = {Proceedings of KDD'12s},
  pages={33--41},
  year={2012},
  organization={ACM}
}

@article{kitsak2010identification,
  title={Identification of influential spreaders in complex networks},
  author={Kitsak, M. and Gallos, L.K. and Havlin, S. and Liljeros, F. and Muchnik, L. and Stanley, H.E. and Makse, H.A.},
  journal={Nature Physics},
  volume={6},
  number={11},
  pages={888--893},
  year={2010}
}

@article{Aad20121,
title = "{Observation of a new particle in the search for the Standard Model Higgs boson with the ATLAS detector at the LHC}",
journal = "Physics Letters B ",
volume = "716",
number = "1",
pages = "1 - 29",
year = "2012",
note = "",
issn = "0370-2693",
author = "{The ATLAS Collaboration}",
}

@article{Chatrchyan201230,
title = "Observation of a new boson at a mass of 125 GeV with the CMS experiment at the LHC",
journal = "Physics Letters B ",
volume = "716",
number = "1",
pages = "30 - 61",
year = "2012",
note = "",
issn = "0370-2693",
author = "{The CMS Collaboration}"
}

@book{Mur12_Machine,
  title={Machine Learning: a Probabilistic Perspective},
  author={Murphy, Kevin P},
  year={2012},
  publisher={MIT press}
}

@book{goodfellow2016deep,
  title={Deep Learning},
  author={Goodfellow, Ian and Bengio, Yoshua and Courville, Aaron},
  year={2016},
  publisher={MIT Press}
}

@article{keeling2005networks,
  title={Networks and epidemic models},
  author={Keeling, M.J. and Eames, K.T.D.},
  journal={Journal of the Royal Society Interface},
  volume={2},
  number={4},
  pages={295--307},
  year={2005}
}

@book{anderson1992infectious,
  address = {Oxford and New York},
  title={Infectious diseases of humans: dynamics and control},
  author = {Anderson, Roy M and May, Robert M},
  volume={28},
  publisher = {Oxford University Press},
  year = {1991}
}

@inproceedings{romero2011differences,
  title={Differences in the mechanics of information diffusion across topics: idioms, political hashtags, and complex contagion on {Twitter}},
  author={Romero, D.M. and Meeder, B. and Kleinberg, J.},
  booktitle={Proceedings of WWW'11},
  pages={695--704},
  year={2011},
  organization={ACM}
}

@inproceedings{kempe2003maximizing,
  title={Maximizing the spread of influence through a social network},
  author={Kempe, D. and Kleinberg, J. and Tardos, {\'E}.},
  booktitle = {Proceedings of KDD'03},
  pages={137--146},
  year={2003},
  organization={ACM}
}

@article{gonzalez2011dynamics,
  title={The dynamics of protest recruitment through an online network},
  author={Gonz{\'a}lez-Bail{\'o}n, S. and Borge-Holthoefer, J. and Rivero, A. and Moreno, Y.},
  journal={Scientific Reports},
  number={197},
  volume={1},
  year={2011}
}

@inproceedings{kwak2010twitter,
  title={{What is Twitter, a social network or a news media?}},
  author={Kwak, H. and Lee, C. and Park, H. and Moon, S.},
  booktitle={Proceedings of WWW'10},
  pages={591--600},
  year={2010}
}

@article{Bass01011969,
author = {Bass, Frank M.}, 
title = {A New Product Growth for Model Consumer Durables},
volume = {15}, 
number = {5}, 
pages = {215-227}, 
year = {1969},
journal = {Management Science} 
}

@book{hidalgo2015information,
  title={Why information grows: The evolution of order, from atoms to economies},
  author={Hidalgo, Cesar},
  year={2015},
  publisher={Basic Books}
}

@inproceedings{adar2005tracking,
  title={Tracking information epidemics in blogspace},
  author={Adar, Eytan and Adamic, Lada A},
  booktitle={Proceedings of the 2005 IEEE/WIC/ACM International Conference on Web Intelligence},
  pages={207--214},
  year={2005},
  organization={IEEE Computer Society}
}

@article{aral2009distinguishing,
  title={Distinguishing influence-based contagion from homophily-driven diffusion in dynamic networks},
  author={Aral, Sinan and Muchnik, Lev and Sundararajan, Arun},
  journal={Proceedings of the National Academy of Sciences},
  volume={106},
  number={51},
  pages={21544--21549},
  year={2009},
  publisher={National Academy of Sciences}
}

@article {Williams160196,
	author = {Williams, Matthew J. and Musolesi, Mirco},
	title = {Spatio-temporal networks: reachability, centrality and robustness},
	volume = {3},
	number = {6},
	year = {2016},
	publisher = {The Royal Society},
	journal = {Royal Society Open Science}
}

@inproceedings{cha2009measurement,
  title={A measurement-driven analysis of information propagation in the {Flickr} social network},
  author={Cha, Meeyoung and Mislove, Alan and Gummadi, Krishna P},
  booktitle={Proceedings of WWW'09},
  pages={721--730},
  year={2009},
  organization={ACM}
}

@inproceedings{gruhl2004information,
  title={Information diffusion through blogspace},
  author={Gruhl, Daniel and Guha, Ramanathan and Liben-Nowell, David and Tomkins, Andrew},
  booktitle={Proceedings of WWW'04},
  pages={491--501},
  year={2004},
  organization={ACM}
}

@article{longley2016geo,
  title={{Geo-temporal Twitter demographics}},
  author={Longley, Paul A and Adnan, Muhammad},
  journal={International Journal of Geographical Information Science},
  volume={30},
  number={2},
  pages={369--389},
  year={2016},
  publisher={Taylor \& Francis}
}

@article{mok2010does,
  title={Does distance matter in the age of the Internet?},
  author={Mok, Diana and Wellman, Barry and Carrasco, Juan},
  journal={Urban Studies},
  volume={47},
  number={13},
  pages={2747--2783},
  year={2010},
  publisher={SAGE Publications}
}

@inproceedings{scellato2010distance,
  title={Distance Matters: Geo-social Metrics for Online Social Networks.},
  author={Scellato, Salvatore and Mascolo, Cecilia and Musolesi, Mirco and Latora, Vito},
  booktitle={Proceedings of WOSN'10},
  year={2010}
}

@book{cairncross2001death,
  title={The death of distance: How the communications revolution is changing our lives},
  author={Cairncross, Frances},
  year={2001},
  publisher={Harvard Business Press}
}

@article{hristova16:multilayer,
  title={A multilayer approach to multiplexity and link prediction in online geo-social networks},
  author={Desislava Hristova and Anastasios Noulas and Chlo{\"e} Brown and Mirco Musolesi and Cecilia Mascolo},
  journal={EPJ Data Science},
  volume={5},
  number={24},
  year={2016},
  publisher={Springer}
}

@inproceedings{hristova16:measuring,
 title={Measuring Urban Social Diversity Using Interconnected Geo-Social Networks},
 author={Desislava Hristova and Matthew J. Williams and Mirco Musolesi and Pietro Panzarasa and Cecilia Mascolo},
 booktitle={Proceedings of WWW'16},
 year={2016}
 }

@article{leetaru2013mapping,
  title={{Mapping the global Twitter heartbeat: The geography of Twitter}},
  author={Leetaru, Kalev and Wang, Shaowen and Cao, Guofeng and Padmanabhan, Anand and Shook, Eric},
  journal={First Monday},
  volume={18},
  number={5},
  year={2013}
}

@article{goodchild2007citizens,
  title={Citizens as sensors: the world of volunteered geography},
  author={Goodchild, Michael F},
  journal={GeoJournal},
  volume={69},
  number={4},
  pages={211--221},
  year={2007},
  publisher={Springer}
}

@book{mitchell2015web,
  title={Web Scraping with Python: Collecting Data from the Modern Web},
  author={Mitchell, Ryan},
  year={2015},
  publisher={O'Reilly Media, Inc.}
}

@book{moretti2005graphs,
  title={Graphs, Maps, Trees: Abstract Models for a Literary History},
  author={Moretti, Franco},
  year={2005},
  publisher={Verso}
}

@book{manning1999foundations,
  title={Foundations of Statistical Natural Language Processing},
  author={Manning, Christopher D and Sch{\"u}tze, Hinrich},
  year={1999},
  publisher={MIT Press}
}

@inproceedings{ahlers2013assessment,
  title={Assessment of the accuracy of GeoNames gazetteer data},
  author={Ahlers, Dirk},
  booktitle={Proceedings of the 7th Workshop on Geographic Information Retrieval},
  pages={74--81},
  year={2013},
  organization={ACM}
}

@book{reas2007processing,
  title={Processing: a Programming Handbook for Visual Designers and Artists},
  author={Reas, Casey and Fry, Ben},
  year={2014},
  publisher={MIT Press}
}

@book{longley2015geographic,
  title={Geographic Information Systems and Science.},
  author={Longley, Paul A and Goodchild, Michael F and Maguire, David J and Rhind, David W and others},
  year={2015},
  publisher={John Wiley \& Sons Ltd}
}

@article{guimera2005worldwide,
  title={The worldwide air transportation network: Anomalous centrality, community structure, and cities' global roles},
  author={Guimera, Roger and Mossa, Stefano and Turtschi, Adrian and Amaral, LA Nunes},
  journal={Proceedings of the National Academy of Sciences},
  volume={102},
  number={22},
  pages={7794--7799},
  year={2005},
  publisher={National Acad Sciences}
}

@article{hawelka2014geo,
  title={{Geo-located Twitter as proxy for global mobility patterns}},
  author={Hawelka, Bartosz and Sitko, Izabela and Beinat, Euro and Sobolevsky, Stanislav and Kazakopoulos, Pavlos and Ratti, Carlo},
  journal={Cartography and Geographic Information Science},
  volume={41},
  number={3},
  pages={260--271},
  year={2014},
  publisher={Taylor \& Francis}
}

@article{musolesi2014big,
  title={Big Mobile Data Mining: Good or Evil?},
  author={Musolesi, Mirco},
  journal={IEEE Internet Computing},
  volume={18},
  number={1},
  pages={78--81},
  year={2014},
  publisher={IEEE}
}

@article{bilke2001topological,
  title={Topological properties of citation and metabolic networks},
  author={Bilke, Sven and Peterson, Carsten},
  journal={Physical Review E},
  volume={64},
  number={3},
  pages={036106},
  year={2001},
  publisher={APS}
}

@article{mislove2011understanding,
  title={{Understanding the Demographics of Twitter Users}},
  author={Mislove, Alan and Lehmann, Sune and Ahn, Yong-Yeol and Onnela, Jukka-Pekka and Rosenquist, J Niels},
  journal={Proceedings of ICWSM'11},
  volume={11},
  pages={5th},
  year={2011}
}

@article{ellison2007benefits,
  title={{"The benefits of Facebook ``friends'': Social capital and college students? use of online social network sites"}},
  author={Ellison, Nicole B and Steinfield, Charles and Lampe, Cliff},
  journal={Journal of Computer-Mediated Communication},
  volume={12},
  number={4},
  pages={1143--1168},
  year={2007},
  publisher={Wiley Online Library}
}

@article{porter2009science,
  title={Is science becoming more interdisciplinary? Measuring and mapping six research fields over time},
  author={Porter, Alan and Rafols, Ismael},
  journal={Scientometrics},
  volume={81},
  number={3},
  pages={719--745},
  year={2009},
  publisher={Akad{\'e}miai Kiad{\'o}, co-published with Springer Science+ Business Media BV, Formerly Kluwer Academic Publishers BV}
}

@article{lehmann2003citation,
  title={Citation networks in high energy physics},
  author={Lehmann, Sune and Lautrup, Benny and Jackson, AD},
  journal={Physical Review E},
  volume={68},
  number={2},
  pages={026113},
  year={2003},
  publisher={APS}
}

@article{gonzalez2013broadcasters,
  title={Broadcasters and hidden influentials in online protest diffusion},
  author={Gonz{\'a}lez-Bail{\'o}n, Sandra and Borge-Holthoefer, Javier and Moreno, Yamir},
  journal={American Behavioral Scientist},
  pages={0002764213479371},
  year={2013},
  publisher={Sage Publications}
}

@article{watts2007influentials,
  title={Influentials, networks, and public opinion formation},
  author={Watts, Duncan J and Dodds, Peter Sheridan},
  journal={Journal of consumer research},
  volume={34},
  pages={441--458},
  year={2007}
}

@article{noulas2012tale,
  title={A tale of many cities: universal patterns in human urban mobility},
  author={Noulas, Anastasios and Scellato, Salvatore and Lambiotte, Renaud and Pontil, Massimiliano and Mascolo, Cecilia},
  journal={PloS ONE},
  volume={7},
  number={5},
  pages={e37027},
  year={2012},
  publisher={Public Library of Science}
}

@inproceedings{tang2010analysing,
  title={Analysing information flows and key mediators through temporal centrality metrics},
  author={Tang, John and Musolesi, Mirco and Mascolo, Cecilia and Latora, Vito and Nicosia, Vincenzo},
  booktitle={Proceedings of WOSN'10},
  pages={3},
  year={2010},
  organization={ACM}
}

@article{tang2010small,
  title={Small-world behavior in time-varying graphs},
  author={Tang, John and Scellato, Salvatore and Musolesi, Mirco and Mascolo, Cecilia and Latora, Vito},
  journal={Physical Review E},
  volume={81},
  number={5},
  pages={055101},
  year={2010},
  publisher={American Physical Society}
}

@article{holme2012temporal,
  title={Temporal networks},
  author={Holme, Petter and Saram{\"a}ki, Jari},
  journal={Physics Reports},
  volume={519},
  number={3},
  pages={97--125},
  year={2012},
  publisher={Elsevier}
}

@inproceedings{lima2012spatial,
  title={Spatial dissemination metrics for location-based social networks},
  author={Lima, Antonio and Musolesi, Mirco},
  booktitle={Proceedings of UbiComp'12},
  pages={972--979},
  year={2012},
  organization={ACM}
}

@inproceedings{Silva14:youare,
 title = {{You are What you Eat (and Drink): Identifying Cultural Boundaries by Analyzing Food \& Drink Habits in Foursquare}},
 author = {Thiago Silva and Pedro {Vaz De Melo} and Jussara Almeida and Mirco Musolesi and Antonio Louriero},
 booktitle = {Proceedings of ICWSM'14},
 address = {Ann Arbor, Michigan, USA},
 month = {June},
 year = {2014},
 }

@book{newman2010networks,
  title={Networks: an Introduction},
  author={Newman, Mark},
  year={2010},
  publisher={Oxford University Press}
}

@inproceedings{laurila2012mobile,
  title={The mobile data challenge: Big data for mobile computing research},
  author={Laurila, Juha K and Gatica-Perez, Daniel and Aad, Imad and Bornet, Olivier and Do, Trinh-Minh-Tri and Dousse, Olivier and Eberle, Julien and Miettinen, Markus and others},
  booktitle={Pervasive Computing '10},
  year={2012}
}

@article{nekovee2007theory,
  title={Theory of rumour spreading in complex social networks},
  author={Nekovee, Maziar and Moreno, Yamir and Bianconi, Ginestra and Marsili, Matteo},
  journal={Physica A: Statistical Mechanics and its Applications},
  volume={374},
  number={1},
  pages={457--470},
  year={2007},
  publisher={Elsevier}
}

@article{lazer2009life,
  title={Life in the network: the coming age of computational social science},
  author={Lazer, David and Pentland, Alex Sandy and Adamic, Lada and Aral, Sinan and Barabasi, Albert Laszlo and Brewer, Devon and Christakis, Nicholas and Contractor, Noshir and Fowler, James and Gutmann, Myron and others},
  journal={Science},
  volume={323},
  number={5915},
  pages={721},
  year={2009}
}

@article{liben2008tracing,
  title={Tracing information flow on a global scale using Internet chain-letter data},
  author={Liben-Nowell, David and Kleinberg, Jon},
  journal={Proceedings of the National Academy of Sciences},
  volume={105},
  number={12},
  pages={4633--4638},
  year={2008},
  publisher={National Acad Sciences}
}

@article{lerman2010information,
  title={{Information contagion: An empirical study of the spread of news on Digg and Twitter social networks}},
  author={Lerman, Kristina and Ghosh, Rumi},
  journal={Proceedings of ICWSM'10},
  volume={10},
  pages={90--97},
  year={2010}
}

@article{boccaletti2006complex,
  title={Complex networks: Structure and dynamics},
  author={Boccaletti, Stefano and Latora, Vito and Moreno, Yamir and Chavez, Martin and Hwang, D-U},
  journal={Physics reports},
  volume={424},
  number={4},
  pages={175--308},
  year={2006},
  publisher={Elsevier}
}

@article{karsai2011small,
  title={Small but slow world: How network topology and burstiness slow down spreading},
  author={Karsai, M{\'a}rton and Kivel{\"a}, Mikko and Pan, Raj Kumar and Kaski, Kimmo and Kert{\'e}sz, J{\'a}nos and Barab{\'a}si, A-L and Saram{\"a}ki, Jari},
  journal={Physical Review E},
  volume={83},
  number={2},
  pages={025102},
  year={2011},
  publisher={APS}
}

@book{barrat2008dynamical,
  title={Dynamical Processes on Complex Networks},
  author={Barrat, Alain and Barthelemy, Marc and Vespignani, Alessandro},
  year={2008},
  publisher={Cambridge University Press}
}

@article{pastor2015epidemic,
  title={Epidemic processes in complex networks},
  author={Pastor-Satorras, Romualdo and Castellano, Claudio and Van Mieghem, Piet and Vespignani, Alessandro},
  journal={Reviews of modern physics},
  volume={87},
  number={3},
  pages={925},
  year={2015},
  publisher={APS}
}

@article{goldenberg2001talk,
  title={Talk of the network: A complex systems look at the underlying process of word-of-mouth},
  author={Goldenberg, Jacob and Libai, Barak and Muller, Eitan},
  journal={Marketing letters},
  volume={12},
  number={3},
  pages={211--223},
  year={2001},
  publisher={Springer}
}

@article{dodds2011temporal,
  title={{Temporal patterns of happiness and information in a global social network: Hedonometrics and Twitter}},
  author={Dodds, Peter Sheridan and Harris, Kameron Decker and Kloumann, Isabel M and Bliss, Catherine A and Danforth, Christopher M},
  journal={PloS ONE},
  volume={6},
  number={12},
  pages={e26752},
  year={2011},
  publisher={Public Library of Science}
}

@inproceedings{brandes2009network,
  title={{Network analysis of collaboration structure in Wikipedia}},
  author={Brandes, Ulrik and Kenis, Patrick and Lerner, J{\"u}rgen and van Raaij, Denise},
  booktitle={Proceedings of WWW'09},
  pages={731--740},
  year={2009},
  organization={ACM}
}

@article{yasseri2012dynamics,
  title={{Dynamics of conflicts in Wikipedia}},
  author={Yasseri, Taha and Sumi, Robert and Rung, Andr{\'a}s and Kornai, Andr{\'a}s and Kert{\'e}sz, J{\'a}nos},
  journal={PloS ONE},
  volume={7},
  number={6},
  pages={e38869},
  year={2012},
  publisher={Public Library of Science}
}

@book{margetts2015political,
  title={Political Turbulence: How Social Media Shape Collective Action},
  author={Margetts, Helen and John, Peter and Hale, Scott and Yasseri, Taha},
  year={2015},
  publisher={Princeton University Press}
}

@inproceedings{hristova2013life,
  title={The Life of the Party: Impact of Social Mapping in OpenStreetMap.},
  author={Hristova, Desislava and Quattrone, Giovanni and Mashhadi, Afra J and Capra, Licia},
  booktitle={Proceedings of ICWSM'13},
  year={2013}
}

@article{davis2013global,
  title={Global spatio-temporal patterns in human migration: a complex network perspective},
  author={Davis, Kyle F and D'Odorico, Paolo and Laio, Francesco and Ridolfi, Luca},
  journal={PLoS ONE},
  volume={8},
  number={1},
  pages={e53723},
  year={2013},
  publisher={Public Library of Science}
}

@article{internetcomputing,
	author={Andrew T. Campbell and Shane B. Eisenman and Nicholas D. Lane and Emiliano Miluzzo and Ronald Peterson and Hong Lu and Xiao Zheng and Mirco Musolesi and Kristof Fodor and Gahng-Seop Ahn},
	title={The Rise of People-Centric Sensing},
	journal={IEEE Internet Computing Special Issue on Mesh Networks},
	month={June/July},
	year={2008}
}

@book{cheshire2014,
  title={London The Information Capital},
  author={Cheshire, James and Uberti, Oliver},
  year={2014},
  publisher={Particular Books}
}

@article{cha2010measuring,
  title={Measuring User Influence in Twitter: The Million Follower Fallacy.},
  author={Cha, Meeyoung and Haddadi, Hamed and Benevenuto, Fabricio and Gummadi, P Krishna},
  journal={Proceedings of ICWSM'10},
  volume={10},
  number={10-17},
  pages={30},
  year={2010}
}

@article{fortunato2018science,
  title={Science of science},
  author={Fortunato, Santo and Bergstrom, Carl T and B{\"o}rner, Katy and Evans, James A and Helbing, Dirk and Milojevi{\'c}, Sta{\v{s}}a and Petersen, Alexander M and Radicchi, Filippo and Sinatra, Roberta and Uzzi, Brian and others},
  journal={Science},
  volume={359},
  number={6379},
  pages={eaao0185},
  year={2018},
  publisher={American Association for the Advancement of Science}
}

@inproceedings{CanMus15_trajectories,
 author = {Luca Canzian and Mirco Musolesi},
 title = {{Trajectories of Depression: Unobtrusive Monitoring of Depressive States by means of Smartphone Mobility Traces Analysis}},
 booktitle = {Proceedings of the 2015 ACM International Joint Conference on Pervasive and Ubiquitous Computing (ACM UbiComp'15)},
 year = {2015},
 month = {September},
 location = {Osaka, Japan},
 publisher = {ACM}
 }

@article{ozkul2013you,
  title={{`You're virtually there': Mobile communication practices, locational information sharing and place attachment}},
  author={{\"O}zkul, Didem},
  journal={First Monday},
  volume={18},
  number={11},
  year={2013}
}

\end{document}